\documentclass[bibyear]{aa} 

\usepackage{graphicx}
\usepackage{txfonts}
\usepackage{natbib}
\bibpunct{(}{)}{;}{a}{}{,} 

\begin{document} 

   \title{Once in a blue stream} 

   \subtitle{Detection of recent star formation in the NGC~7241 stellar stream with MEGARA}

   \author{David Mart{\'\i}nez-Delgado\inst{1}\fnmsep\thanks{Talentia Senior Fellow, dmartinez@iaa.es}
   \and
Santi Roca-F\`abrega\inst{2,3}\fnmsep\thanks{sroca01@ucm.es}
\and
Armando Gil de Paz\inst{3,4}
\and
Denis Erkal\inst{5}
\and
Juan Mir\'o-Carretero\inst{3}
\and
Dmitry Makarov\inst{6}
\and
Karina T. Voggel\inst{7}
\and
Ryan Leaman\inst{8}
\and
Walter Boschin\inst{9}
\and
Sarah Pearson\inst{10}
\and
Giuseppe Donatiello\inst{11}
\and
Evgenii Rubtsov\inst{12}
\and
Mohammad Akhlaghi\inst{13}
\and
M. Angeles Gomez-Flechoso\inst{3,4}
\and
Samane Raji\inst{14}
\and
Dustin Lang\inst{15}
\and
Adam Block\inst{16}
\and
Jesus Gallego\inst{3,4}
\and
Esperanza Carrasco\inst{17}
\and
Mar\'\i a Luisa Garc\'\i a-Vargas\inst{18}
\and
Jorge Iglesias-P\'{a}ramo\inst{1}
\and
Sergio Pascual\inst{3,4}
\and
Nicolas Cardiel\inst{3,4}
\and
Ana Pérez-Calpena\inst{18}
\and
Africa Castillo-Morales\inst{3,4}
\and
Pedro Gómez-Alvarez\inst{18}}

\institute{Instituto de Astrof\'isica de Andaluc\'ia, CSIC, Glorieta de la Astronom\'\i a, E-18080, Granada, Spain
\and
Lund Observatory, Division of Astrophysics, Department of Physics, Lund University, Box 43, SE-221 00 Lund, Sweden
\and
Departamento de F{\'\i}sica de la Tierra y Astrof{\'\i}sica, Universidad Complutense de Madrid, E-28040 Madrid, Spain
\and
Instituto de Física de Partículas y del Cosmos (IPARCOS), Fac. CC. Físicas, Universidad Complutense de Madrid, Plaza de las Ciencias, 1, E-28040 Madrid, Spain
\and
Department of Physics, University of Surrey, Guildford GU2, 7XH, UK
\and
Special Astrophysical Observatory, Russian Academy of Sciences, Nizhnii Arkhyz, 369167 Russia
\and
Universite de Strasbourg, CNRS, Observatoire astronomique de Strasbourg, UMR 7550, 67000 Strasbourg, France
\and
Department of Astrophysics, University of Vienna, T\"{u}rkenschanzstra$\beta$e 17, 1180 Vienna, Austria
\and
Fundaci\'on G. Galilei - INAF (TNG), Rambla J. A. Fern\'andez P\'erez 7, E-38712 Bre\~na Baja (La Palma), Spain
\and
Center for Cosmology and Particle Physics, Department of Physics, NYU, 726 Broadway, New York, NY 10003, USA
\and
UAI -- Unione Astrofili Italiani /P.I. Sezione Nazionale di Ricerca Profondo Cielo, 72024 Oria, Italy
\and
Sternberg Astronomical Inst., M.V. Lomonosov Moscow State University, Universitetsky prospect 13, Moscow, 119234, Russia
\and
Centro de Estudios de Física del Cosmos de Aragón (CEFCA), Unidad Asociada al CSIC, Plaza San Juan 1, 44001 Teruel, Spain
\and
Department of Physics, Yazd University, University Blvd, Safayieh, Yazd, Iran
\and
Perimeter Institute for Theoretical Physics, 31 Caroline St N, Waterloo, Canada
\and
Steward Observatory, Department of Astronomy, University of Arizona, 933 N. Cherry Avenue, Tucson, AZ 85748, USA
\and
Instituto Nacional de Astrofísica, Óptica y Electrónica, Calle Luis Enrique Erro 1, Tonantzintla, Puebla, Mexico 
\and
FRACTAL S.L.N.E. Calle Tulipán 2, portal 13, 1A, E-28231 Las Rozas de Madrid, Spain
}

  \abstract
   {} 
  {In this work we aim to study the striking case of a narrow blue stream with a possible globular cluster-like progenitor around the NGC~7241 galaxy and its foreground dwarf companion. We want to figure out if the stream was generated by tidal interaction with NGC~7241 or it first interacted with the foreground dwarf companion and later both fell together towards NGC~7241.} 
   {We use four sets of observations, including a follow-up spectroscopic study of this stream based on data taken with the MEGARA instrument at the 10.4-m Gran Telescopio Canarias using the integral field spectroscopy mode, Mount Lemmon 0.80-meter telescope, Telescopio Nazionale Galileo, DESI imaging Legacy surveys and GALEX archival data. We also use high resolution zoom-in cosmological simulations.}
   {Our data suggest that the compact object we detected in the stream is a foreground Milky Way halo star. Near this compact object we detect emission lines overlapping a less compact, bluer, and fainter blob of the stream that is clearly visible in both ultra-violet and optical deep images. From its heliocentric systemic radial velocity  derived from the [\ion{O}{iii}]$\lambda$5007\AA\ lines ($V_{\rm syst}$= 1548.58$\pm$1.80\,km\,s$^{-1}$) and new UV and optical broad-band photometry, we conclude that this over-density could be the actual core of the stream, with an absolute magnitude of M${g}\sim$ -10 and a $g-r$ = 0.08 $\pm$ 0.11, consistent with a remnant of a low-mass dwarf satellite undergoing a current episode of star formation. From the width of the stream and assuming a circular orbit, we calculate that the progenitor mass can be the typical of a dwarf galaxy, but it could also be substantially lower if the stream is on a very radial orbit or it was created by tidal interaction with the companion dwarf instead of with NGC~7241. These estimates also suggest that this is one of the lowest mass streams detected so far beyond the Local Group. Finally, we find that blue stellar streams containing star formation regions are commonly predicted by high-resolution cosmological simulations of galaxies lighter than the Milky Way. This scenario is consistent with the processes explaining the bursty star formation history of some dwarf satellites, which are followed by a gas depletion and a fast quenching once they enter within the virial radius of their host galaxies for the first time. Thus, it is likely that the stream's progenitor is suffering a star-formation burst comparable to those that have shaped the star-formation history of several Local Group dwarfs in the last few Gigayears.}
   {}


   \maketitle
%
\section{Introduction}

Star formation in the low-z universe mostly occurs inside galactic disks through dynamical instabilities \citep{Kennicutt1998,Leroy2008}. These processes dominate the current Star Formation Rate (SFR) in the Universe but are not exclusive. Many authors report observations of star forming regions outside galactic disks \citep[e.g., ][]{ODea2004,Sengupta2009,Howk2018a,Smith2010}. In these regions, a gas shell, a gas stream, or a gas cloud suffers a strong perturbation, collapses and generates a new population of stars. Some examples of this process are the star formation clumps observed in boundaries of radio jets and lobes \citep{Graham1998,Mould2000,ODea2004}, in clouds of extraplanar gas originated by the galactic fountain \citep{Tullmann2003,Howk2018a},  or in highly perturbed regions like the tidal debris left from first interactions during wet major mergers \citep{Sengupta2009,Hibbard2005}. If massive enough this new self-gravitating stellar population lead to the formation of a young star cluster \citep{Howk2018b}, or, in strong tidal interactions, new objects like Tidal Dwarf Galaxies, TDG \citep{DucMirabel1998,Duc2000,Smith2007,Smith2010}. 

Two ingredients are needed in these processes to allow the formation of a new stellar population: the presence of a cold/molecular gas cloud and a strong perturbation. Strong perturbations are present in most galactic systems in many forms, from gas sound waves (e.g., SNe shells or AGN jets) to tidal perturbations by companions or non-axisymmetric structures inside galaxies \citep{Buta2019}. Besides these special conditions, the formation of cold/molecular clouds is not seen outside the disk plane. In the disk outskirts, the dynamics of gas is tightly bounded with the  Circum-Galactic Medium (CGM) properties. High-mass galaxies develop a warm-hot CGM that inhibits the formation of cold clouds of molecular gas in hydrostatic equilibrium far from the disk region \citep{Keres2005}. On the other hand, low-mass galactic systems present the prototypical galactic fountain where metal-rich gas is released by SNe winds to the CGM, or even the Inter-Galactic Medium (IGM), and quickly cools down and falls back to the disk from large radii. So, there is a strong dependence on the distribution of extrapalanar star formation with galaxy virial mass (i.e., CGM's virial temperature) \citep{Fraternali2017}. Similarly, it is favored in low-mass systems that cold gas infalling from the IGM through cold flows, or inside gas rich satellites, penetrates down to the galactic disk, and/or condensates inside gas-rich filaments or tidal structures \citep{Fraternali2017}.

The star formation in the tidal tails generated around in-falling satellites during gas rich minor mergers has only been studied in a few works using Smoothed-Particles Hydrodynamic (SPH) simulations \citep{Kapferer2009}. Current studies of extraplanar star formation have mainly focused on the consequences of major mergers and interactions within massive clusters of galaxies \citep{Tullmann2003,ODea04,Leroy08,Werk08,McDonald12}. Furthermore, predictions from the theory of galaxy formation and evolution points towards a relation between the minimum radius reached by gas inside infalling satellites before being ram pressure stripped, and the properties of the gas surrounding the central galaxy \citep{Hester06}, that are tightly linked with the total mass of the central galaxy \citep{BirnboimDekel03}. If the central galaxy is a low-mass system, i.e., if it has not developed a dense warm-hot CGM, the satellite can reach lower radii before it is totally ram-pressure stripped. In this situation the satellite's gas can shock with the central galaxy HI (from the galactic fountain or inside the disk) and produce extraplanar star formation \citep{Norena2019}.  In this situation, newborn stars will follow a similar orbit as the infalling galaxy \citep{Lim2018}, and, as a result, we would observe a blue stellar stream. 

\begin{figure*}
	\includegraphics[width=0.85\textwidth]{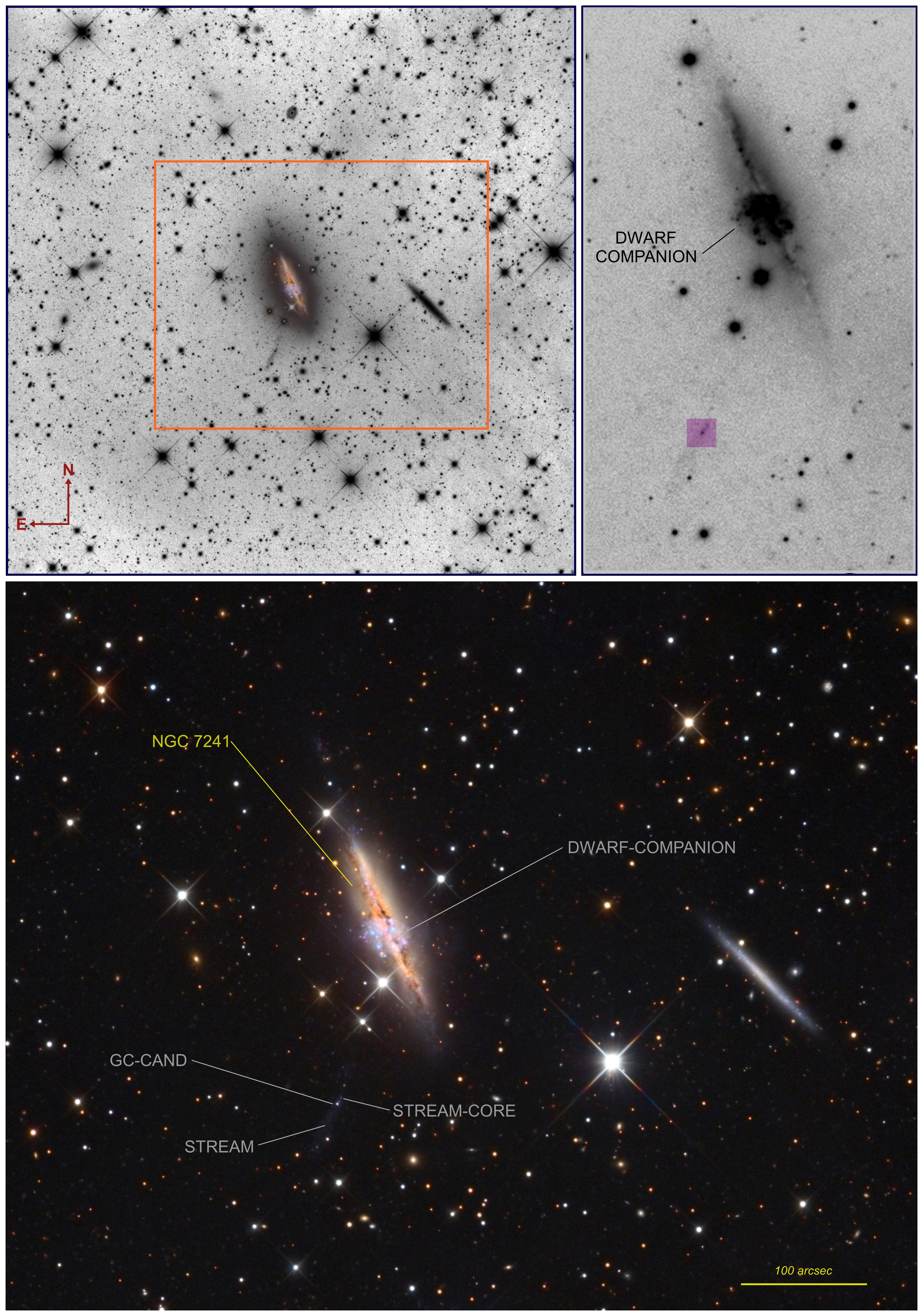}
    \caption{NGC~7241 images including the  blue stream and the dwarf companion projected on its center. {\it Upper-left}: L-filter wide-field image obtained with the Mount Lemmon 0.80-meter $f/7$ Ritchey–Chrétien telescope. The total field-of-view is 22.5 $\times$ 22.5 arcmin. The red square marks the field of the full color image displayed in the bottom panel. {\it Upper-right}: u'-band observations obtained with DOLoRes at the TNG 3.58-meter in La Palma observatory. The purple filled square marks the field of view of the MEGARA IFU field which is placed on the compact object in the stream. {\it Bottom panel}: A zoomed, full color version of the deep image obtained with the Mount Lemmon  0.80-meter telescope, including the NGC~7241 central region marked with a red square in the upper left panel.}
    \label{fig:1}
\end{figure*}

NGC~7241 is a galaxy that is transitioning from the low-mass, with no warm-hot corona, to the high-mass scenario (like the MW), with a spatially extended HI gas disk \citep{Leaman2015}. The first inspections of this galactic system in the UV images from Galex seemed to show that this edge-on disk galaxy have a strong off-plane star formation, although this galaxy shows no morphological evidence of a recent interaction triggering it. The lack of morphological perturbations was also in contradiction with the HI observations by \citet{Giovanelli1984} that showed evidence of the presence of a possible dwarf companion in the field. More recently, \citet{Leaman2015} confirmed the presence of this companion, it is an actively star forming dwarf in the foreground of the NGC~7241 galactic disk. The foreground dwarf hosts the strong off-plane star formation observed in the Galex images. \citet{Leaman2015} also discovered a nearby stellar tidal stream, and, using H$_{\alpha}$, they studied the kinematics of all components in the system and concluded that the stream is associated with the star-forming companion. However, the analysis of the connection between the so-called "companion" and the NGC~7241 was not conclusive. The authors also studied the companion's SFR and observed that it is one order of magnitude higher than expected in a normal galaxy of its mass. These results, in addition to the absence of strong morphology perturbations and of an enhanced SF in the NGC~7241 disk, pointed towards an unknown but potentially complex interaction history. Particularly, it is under study if the infall of a small galactic system towards the companion could create the observed narrow tidal stream.

In this paper, we present follow-up spectroscopic observation of the blue stream of NGC~7241, with the aim of constraining the nature of a compact object found inside the stream. In addition to undertaking a new photometry and structural study of this stream,  we explore the formation scenario of this rare blue streams in the nearby Universe using cosmological simulations. In this context, we explore if our observations of a possible star forming progenitor are showing a process similar to the one which shaped the SFH of some dwarfs in the Local Group that show an early star formation peak followed by a fast quenching \citep{Weisz2015}.

\section{OBSERVATIONS AND DATA REDUCTION}

\subsection{Gran Telescopio de Canarias MEGARA spectroscopy observations}
\label{sec:megara} 

The brightest region of the NGC~7241 stream was observed with the {\it Multi Espectr\'ografo en GTC de Alta Resoluci\'on para Astronom\'\i a} (MEGARA) instrument at the Gran Telescopio Canarias (GTC) on the night of June 6$^{\mathrm{th}}$ 2020 using the Integral Field Unit (IFU) in the LR-B spectral configuration with a power resolution of R$\sim$5000 in the wavelength interval 4332 -- 5199 \r{A}, for a total exposure time of 3600\,s. The target field, with a field-of-view of 12.5$\arcsec$ $\times $ 11.3$\arcsec$ (purple filled square in Fig.~\ref{fig:1} upper right panel),  was centered on the position of the blue compact source embedded in the stream (named {\it $GC-CAND$} hereafter). This observing time was split into three consecutive exposures of 1200\,s to allow for cosmic-ray removal. The three individual exposures were processed using the MEGARA Data Reduction Pipeline version 0.10.1 \citep[{\sc megaradrp}; see][]{2018zndo...2206856P} and combined to generate the final Row-Stacked-Spectra (RSS hereafter) FITS frame that was used for our analysis. 

The data processing included bias subtraction, gain normalization of the two amplifiers, tracing and extraction of the 567 object plus 56 sky fiber spectra, wavelength calibration, flat-fielding correction of the blue-to-red and fiber-to-fiber response and absolute flux calibration. The fiber tracing, extraction modeling and flat-fielding ({\sc TraceMap}, {\sc ModelMap} and {\sc FiberFlat} recipes, respectively) were performed using halogen lamp observations taken at the end of night. The wavelength calibration was performed using observations of the 5 ThAr lamps available at the Instrument Calibration Module (ICM). The standard deviation of our wavelength calibration solution achieved was 0.021\,\AA\ for a reciprocal linear dispersion of 0.22\,\AA\,pix$^{-1}$. In order to perform absolute calibration of these data the spectrophotometric standard star HR~5501 was also observed during that night and processed using the {\sc megaradrp}. More information on the processing of MEGARA data using the {\sc megaradrp} can be found in \cite{2020MNRAS.493..871G}.

\subsection{Mount Lemmon 0.80-meter imaging observations}\label{sec:2.2}

 We collected deep imaging of the field around NGC~7241 at the Mount Lemmon Sky Center (Steward Observatory, University of Arizona) with an 80 cm aperture $f/7$ Ritchey–Chrétien telescope. We used a SBIG STX16803 CCD camera was that provided a pixel scale of 0.33$\arcsec$\,pixels$^{-1}$ over a $22.5\arcmin \times 22.5\arcmin$ field of view. We obtained a set of 32 individual 1200 second  images with an Astrodon Gen2 Tru-Balance E-series luminance filter over several photometric nights between 2017 June and 2017 July. We obtained this data by remote observations. In the upper left panel of Fig.~\ref{fig:1} we show the image we took using the luminance filter that is a wide-band, nearly top-hat filter, that transmits from $400\lesssim \lambda \ (\mathrm{nm})\lesssim 700$, and broadly covers the more familiar $g$ and $r$ bands. Each individual exposure was reduced following standard image processing procedures for dark subtraction, bias correction, and flat fielding adopted for the larger stream survey \citep{2010AJ....140..962M}. The images were combined to create a final co-added image with a total exposure time of 38400\,s.

\subsection{Telescopio Nazionale Galileo imaging observations}

We took deep images of the field of NGC~7241 with the instrument DOLoRes (Device Optimized for the Low Resolution) of the 3.58m Italian Telescopio Nazionale Galileo (TNG, Roque de Los Muchachos Observatory, La Palma, Spain) on December 10, 2018. These observations include 9 x 300s exposures in the u'-band (binning 2x2, pixel scale 0.504") with a field-of-view of 8.6$\arcmin$'x8.6 $\arcmin$ and a seeing measured in this band of $\sim$1.3" (see Fig.~\ref{fig:1}, upper right panel).

The pre-processing of the data was performed in a standard way using IRAF tasks, i.e. dividing the trimmed and bias subtracted frames by a master flat field produced from multiple twilight sky-flat exposures. Then, a final image was obtained by performing a median stack of the preprocessed images. In the last stage, the astrometric
calibration of this image was performed using the astrometry.net tool \citep{lang2010}.

\subsection{DESI Imaging Legacy surveys and GALEX archival data}\label{sec:legacypip}

The DESI Legacy Imaging Surveys compile optical data in three optical bands ($g$, $r,$ and $z$) obtained by three different imaging projects: The DECam Legacy Survey (DECaLS), the Beijing-Arizona Sky Survey (BASS), and the Mayall $z$-band Legacy Survey (MzLS) \citep{2019ApJS..245....4Z, 2019AJ....157..168D}. The DESI Legacy Imaging Survey data releases also include re-reduced public DECam data from the DES \citep[][]{2018ApJS..239...18A}. An image cutout centred on NGC~7241 were subsequently obtained by coadding images of these systems taken by the DES \cite[]{DES16} using the DECam. These data were reprocessed using the \textsc{legacypipe} software of the DESI Legacy imaging surveys \citep[see e.g. Fig. 2 in ][]{2021arXiv210406071M}. In short, each image including our target galaxy is astrometrically calibrated to Gaia-DR2 and photometrically calibrated to the Pan-STARRS PS1 survey, and subsequently resampled to a common pixel grid and summed with inverse-variance weighting.  Fig.\ref{fig:2} (upper panel) shows the resulting coadded image cutout of the NGC~7241 stream galaxy, which  was used to derive the photometry of stream in the $g$, $r$, and $z$-bands (see Sec.\ref{sec:3.2}).

The region around NGC~7241 was also observed by the Galaxy Evolution
Explorer \citep[GALEX; ][]{2005ApJ...619L...1M}  on August 23th 2003 simultaneuosly in both its FUV and NUV bands as part of the GALEX All-sky Imaging Survey (AIS). The valid exposure times acquired on this field were 120 seconds in the FUV channel and 175 seconds in the NUV channel.

\section{RESULTS}

\subsection{Radial velocity from MEGARA observations}\label{sec:radvel}

\begin{figure}
	\includegraphics[width=\columnwidth]{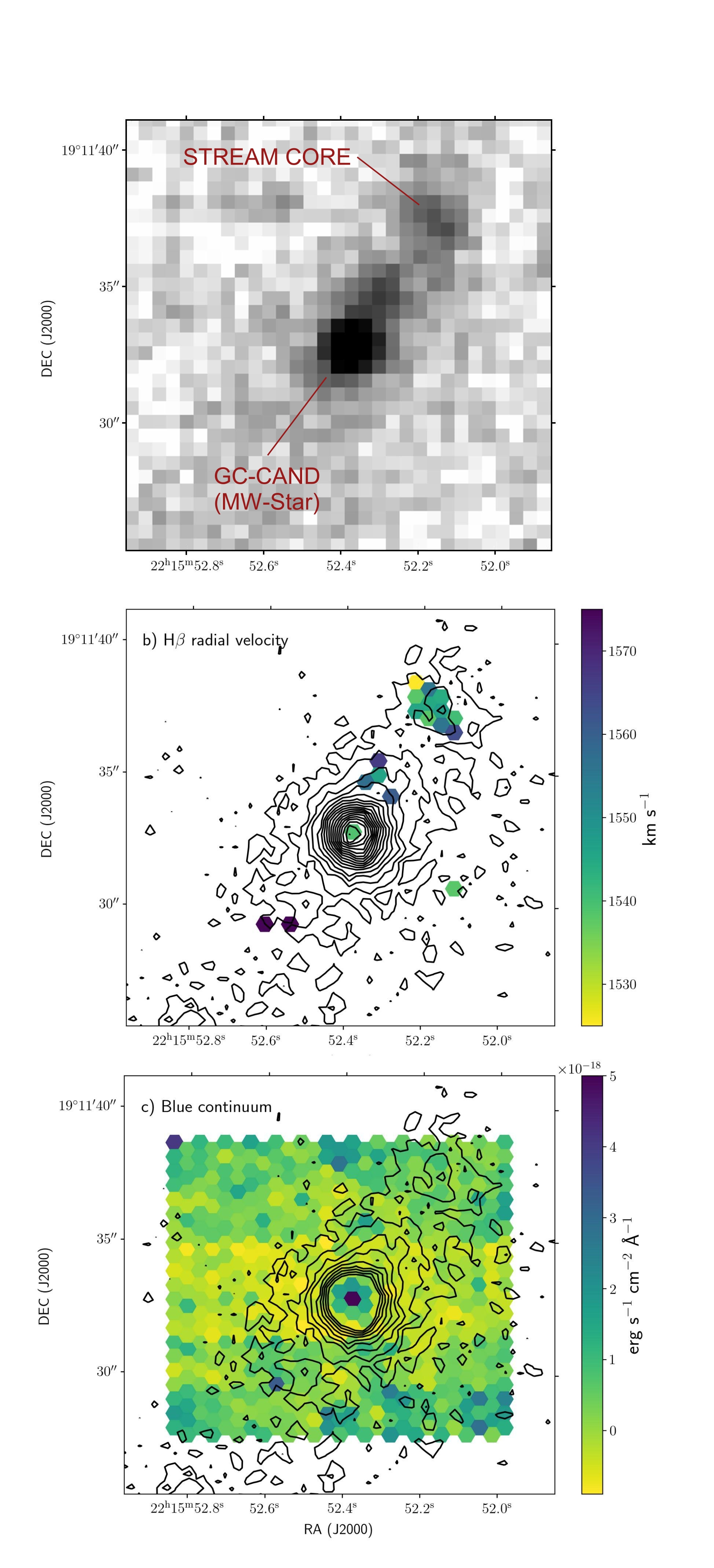}
    \caption{({\bf a)} A zoomed region in the position of the stream taken from the TNG $u'$-band image. It is centered in the compact object targeted in our campaign as possible progenitor of the stream. The image also shows a second fainter over-density at $\sim$ 5 arcsec North-West (top-right) labeled as the "stream core";  b) Heliocentric radial velocity obtained from the emission in H$\beta$ line detected in the MEGARA LR-B data. We observe two emission sources, one compact in the center and a second one at the top-right corner. The $g$-band DESI Imaging Legacy data contours are over-plotted. The $g$-band contours shown correspond to the range between 28 and 31\,AB\,mag\,arcsec$^{-2}$ and are equally spaced in linear (flux) scale.   {\bf c)} Continuum image of the stream nucleus candidate obtained after averaging the flux from the MEGARA LR-B datacube in the (rest-frame) wavelength range between 480-484\,nm and 487-495\,nm. The same $g$-band contours as those used for panel b, except for those brighter than 28.7\,AB\,mag, are plotted here.}
    \label{fig:2}
\end{figure}

The blue stream of NGC~7241 (see Fig.~\ref{fig:1}, bottom panel) is a $\sim$45" long tail that seems to emanate from a bright compact object ({\it GC-CAND}) located at a position angle of $\sim$150 degrees (measured from North to East). The visual inspection of the  Row-Stacked Spectrum (RSS hereafter) 2D FITS file created by the {\sc megaradrp}\footnote{More information on the processing of MEGARA data using the {\sc megaradrp} can be found in \cite{2020MNRAS.493..871G}.} from the observation centered at the {\it GC-CAND} (see Sec.~\ref{sec:megara}) reveals the presence of two faint emission lines in several spaxels that correspond to H$\beta$ and [\ion{O}{iii}]$\lambda$5007\AA\ at the approximate redshift of NGC~7241 \citep[z$\sim$0.0048 or 1447$\pm$1\,km\,s$^{-1}$;][]{1993ApJS...88..383L}. Since the location of these lines does not coincide with the brightest spaxels in the continuum, we decided to determine the heliocentric radial velocity map in the vicinity of the possible nucleus of the NGC~7241 stream using these data.  For that purpose, we made use of the analysis and visualization tools distributed along with the {\sc megaradrp}, namely the {\it analyze$_{rss}$} and {\it visualization} codes \citep[see][]{africa_castillo_morales_2020_3932063}. The first code was used to measure the intensity and recession velocity of every emission line detected in any spaxel with a line peak signal-to-noise ratio (S/N) above a given threshold  along with the corresponding errors\footnote{The errors in radial velocity were obtained from the covariance matrix as part of the analysis performed by the {\sc lmfit} python package included in {\sc megaratools} within the {\sc megaradrp} using a Levenberg-Marquardt least-squares minimization method.}. Interestingly, in our case we found a coherent structure with emission above S/N=5 in 11(5) spaxels in the H$\beta$ ([\ion{O}{iii}]$\lambda$5007\AA) line but not at the position of the {\it GC-CAND} (see Fig.~\ref{fig:NBurstsFit}). This emission emanates from a second, fainter over-density just North of the {\it GC-CAND}, which is visible in both $u$'-band TNG image (marked in Fig.~\ref{fig:2}, upper panel) and DES images (marked with a red circle in Fig.~\ref{fig:3}, bottom panel). Following discussion on the actual progenitor of the NGC~7241 stream (see Sec.\ref{sec:progenitor}), we named this feature as {\it STREAM-CORE}. From the spectrum of these 11 spaxels we obtained a observed (topocentric) wavelength for the bluest of the two lines detected of 4885.987$\pm$0.027\,\AA, that corresponds to a heliocentric radial velocity for the {\it STREAM-CORE} of 1547.92$\pm$1.65\,km\,s$^{-1}$ if we assume that this line is H$\beta$ given that the topocentric to heliocentric correction is $+$24.25\,km\,s$^{-1}$. Adding up the flux from all these spaxels, the best-fitting observed wavelength for the reddest line detected is 5032.301$\pm$0.030\,\AA\ or a heliocentric radial velocity of 1547.59$\pm$1.80\,km\,s$^{-1}$ assuming this corresponds to [\ion{O}{iii}]$\lambda$5007\AA. The results of the best fits to both lines are shown in Fig.~\ref{fig:NBurstsFit}. Note that the standard deviation of our wavelength calibration solution is 1.1\,km\,s$^{-1}$ while the reciprocal dispersion yields $\sim$12\,km\,s$^{-1}$ at these wavelengths.

The only absorption or emission detail visible in the spectrum obtained at the {\it GC-CAND} location is at 4857.0~\AA, with a significance of about 5--6 sigma (see Fig.\ref{fig:NBurstsFit}, bottom panel). Assuming the absorption is H\,$\beta$ we see that it is blue shifted by $-269$~km\,s$^{-1}$. To determine the radial velocity from this line we needed to increase the $S/N$ that was originally of $\sim2$. To do that we reduced the resolution from $R\sim5000$ to $R\sim2500$ using Gaussian smoothing and rebinned the spectrum to the scale of 2 pix per FWHM. With these operations we achieve a $S/N\sim9$. Finally, we fitted the resulting spectrum by the {\sc PEGASE.HR} stellar population models~\citep{2004A&A...425..881L} using the \textsc{NBursts} software~\citep{NBursts2007}. The resulting radial velocity for the {\it GC-CAND} is of $V_h=-233\pm33$~km\,s$^{-1}$ with respect to the solar system barycenter.\footnote{We have found that the feature resembling a P-Cygni absorption component is due to the presence of a sky subtraction residuals that can be seen in the original high-resolution spectrum of STREAM-CORE near Hbeta at around 4878 \AA. This feature does not affect the determination of the radial velocity, line flux or line rations in STREAM-CORE given that it is well separated from our line of interest. Note that not only this feature is out of the line window but was also excluded from the continuum windows used to measure the continuum level and the continuum rms (needed to compute the errors in the line flux)
} 

\subsection{Photometry and structural properties}\label{sec:3.2}

The photometry of the NGC~7241 stream in the DES images was carried out using {\it GNU Astronomy Utilities} (Gnuastro)\footnote{\url{http://www.gnu.org/software/gnuastro}}. All the measurements have been done using Gnuastro's {\sc MakeCatalog} on the sky-subtracted image generated by Gnuastro's {\sc NoiseChisel} \citep{Akhlaghi15,Akhlaghi19}. This tool has been developed with emphasis on the detection of low surface brightness, diffuse sources.
Surface brightness limit of the images, giving the depth of the image, were calculated following the standard method of \cite{Roman2020}, i.e. the $3\sigma$ measured value on a 100 arcsec$^2$ aperture and yield 28.99 [mag\,arcsec$^{-2}$] for $g$, 28.42 [mag arcsec$^{-2}$] for $r$ and 28.28 [mag arcsec$^{-2}$] for $u$. 

\begin{figure}
	\includegraphics[width=\columnwidth]{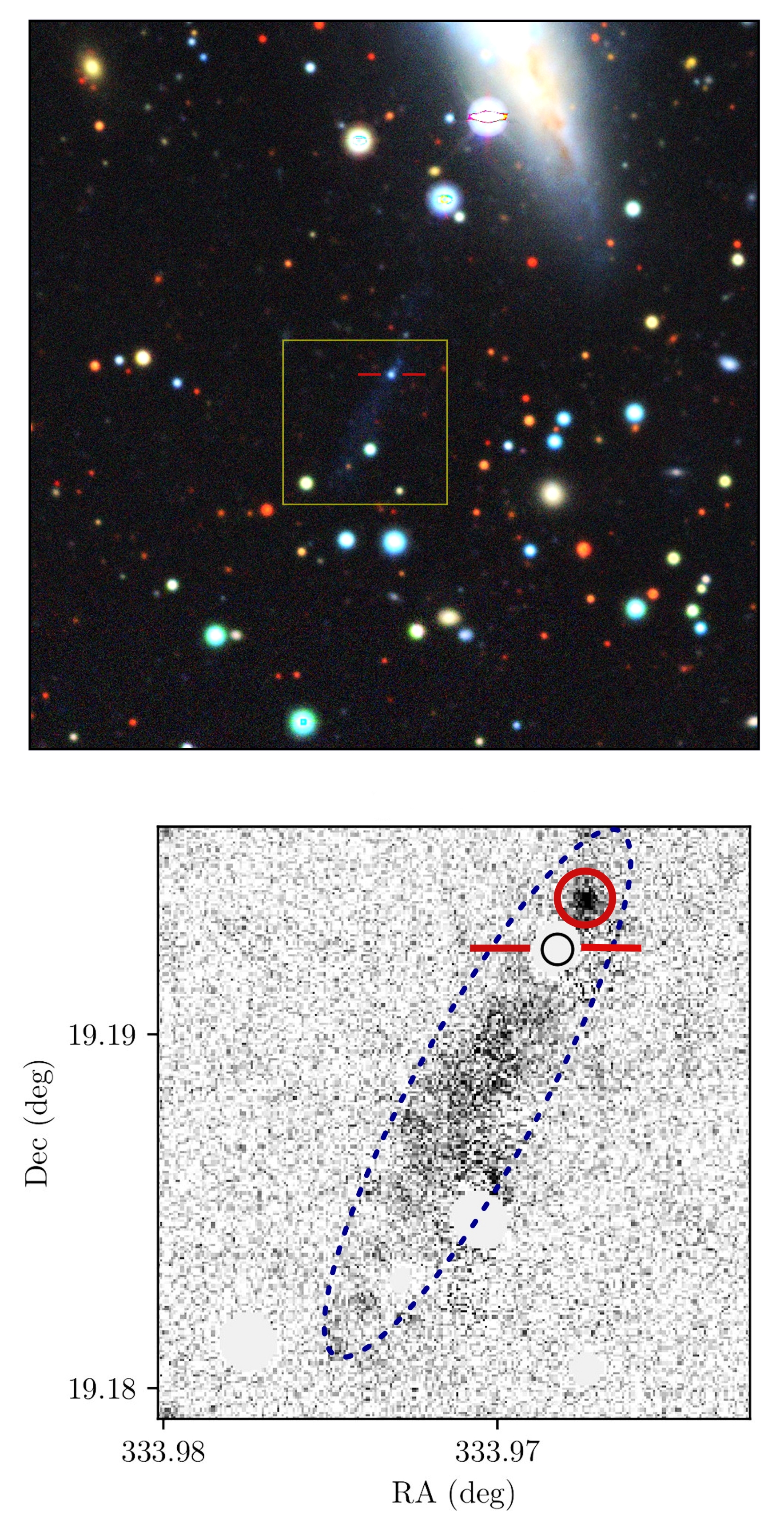}
    \caption{{\it Upper panel}: Image cutout obtained from the DESI Legacy surveys data with {\sc legacypipe} centered in the compact object {\it GC-CAND} (marked with two red lines) embedded in the NGC~7241 stream and classified as a foreground halo star in our study (see Sec.~\ref{sec:progenitor}) . The total field-of-view of this image is 5 $\times$ 5 arcmin. The orange square indicates the extend of the figure shown in the bottom panel. {\it Bottom panel}:  The apertures used to perform the photometry of the stream ({\it dashed blue line}) and its possible progenitor {\it STREAM-CORE} ({\it red circle}) are overlapped in a $r$-band DESI LS image of the stream. The source {\it GC-CAND} (marked with a {\it black circle} encompassed in two red lines) was also masked for deriving the results given in Tab.~\ref{tab:2}.}
    \label{fig:3}
\end{figure}

\begin{figure*}
\centering
\includegraphics[width=\textwidth,angle=0]{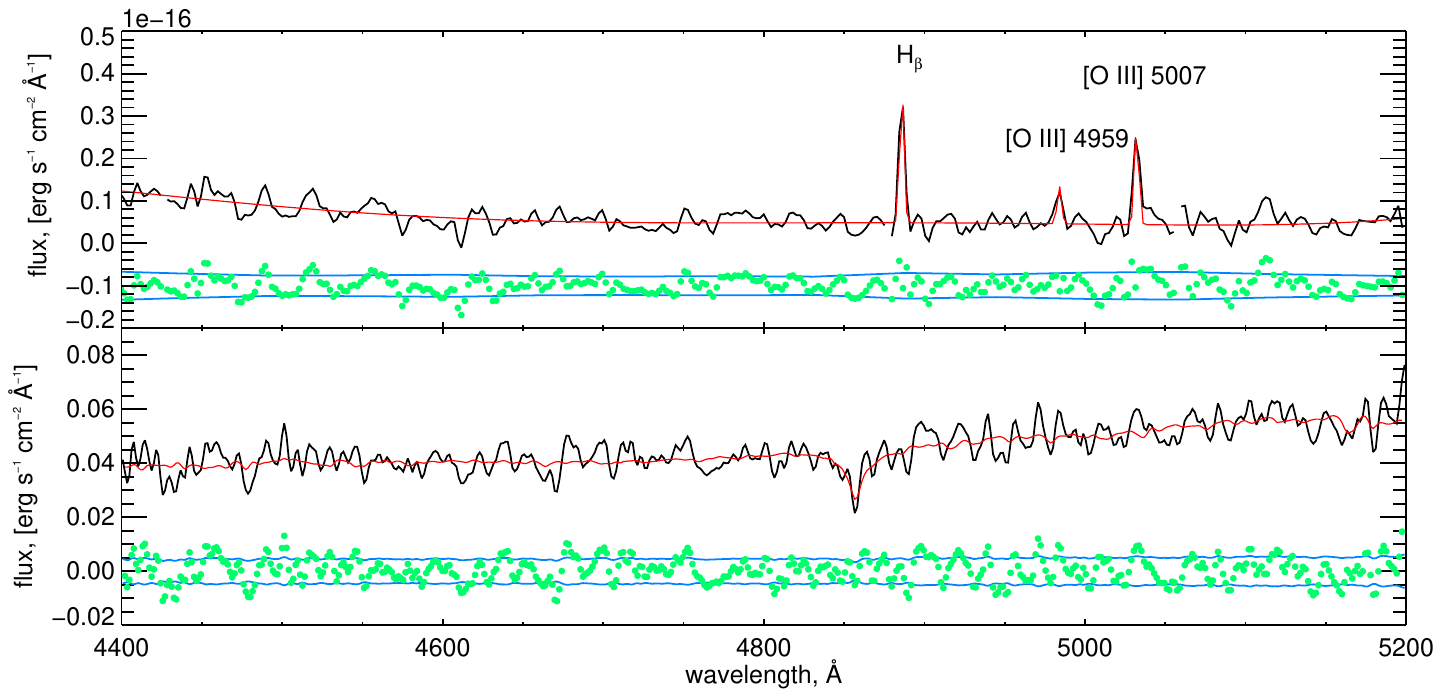}
\caption{{\it Top panel:} MEGARA LR-B spectrum of the {\bf North-West region showing line emission (see top-right region of panel b in Fig.~\ref{fig:2})} fitted by a multi-Gaussian for emission lines with an additive polynomial for continuum. The H$\beta$ line and the [\ion{O}{iii}]$\lambda\lambda$4959,~5007\,\AA\ doublet are identified along with their position for a redshift z=0.00515. {\it Bottom panel:} The spectrum of the compact source embedded in the stream fitted by the \textsc{PEGASE.HR} stellar population models using the \textsc{NBursts} program. The black line represents the smoothed spectrum of the object and the red line corresponds to the fit. The blue lines and green dots show the flux errors and the data residuals (shifted to -0.1 for top panel).}
\label{fig:NBurstsFit}
\end{figure*}

 We used {\sc NoiseChisel} (Gnuastro) to perform sky-subtraction on the DESI images previously processed with the \textsc{legacypipe} (see Sec.\ref{sec:legacypip}). For that purpose, the image is tessellated, with tiles having a configurable number of pixels (typically 40x40). The sky background in the tiles with detection will be obtained by interpolation of the signal in the neighbouring tiles with no detection, and then subtracted locally. In this way the environment of the stream is taken into consideration for the calculation of the sky background. The photometry is then measured on the image once the sky background has been subtracted.
The magnitude and the surface brightness (SB)  have been measured for the $r$ and $g$ bands together with their color $(g-r)$ for the stream  and {\it STREAM-CORE} (marked with a red circle in Fig.~\ref{fig:3}, bottom panel). The results are given in Tab.~\ref{tab:1}. The measurement for the stream was carried out on an elliptical aperture encompassing most of the stream (plotted in Fig.\ref{fig:3} bottom panel with a dashed line). We masked foreground stars and background objects, which includes the star rejected as nucleus of the stream ({\it GC-CAND}). Measurements were carried out with and without masking the compact structure (corresponding to the rows \textit{stream} and \textit{stream+core}, respectively, in Tab.~\ref{tab:1}. This table also shows the measurements of the possible progenitor of the stream itself (labelled as {\it STREAM-CORE}), showing that it is bluer than the stream itself. In order to verify that the ellipse was a good enough fit to the stream contour, and the measurements were thus correct, the surface brightness was also measured on circular apertures placed carefully on the stream to ensure they were completely overlapping with it. The results were within 0.04 and 0.11 [mag arcsec$^{-2}$] of the measurements on the elliptical aperture for $g$ and $r$, respectively, which shows that our results are not depending on the exact chosen aperture.

\begin{table*}
\centering                          
\caption{Photometry results for the NGC~7241 stream from Gnuastro. We include the data from the stream, the structure we labeled {\it STREAM-CORE}, and the combined (TOTAL).} 
\label{tab:1}     
\begin{tabular}{c c c c c c c c c c}        
\hline\hline                 
Source & area     & area     & \textit{m}$_\textrm{g}$ & \textit{m}$_\textrm{r}$ & \textit{m}$_\textrm{u}$ & <$\mu_{g}$> & <$\mu_{r}$> & <$\mu_{u}$> & <\textit{g}-\textit{r}$>_\textrm{stream}$ \\
       & [pixels] & [as$^{2}$] & & [mag] &  &  & [mag as$^{-2}$] &  & [mag] \\
\hline                        
 STREAM            & 7037 & 483.05 & 19.23 $\pm$ 0.01 & 19.07 $\pm$ 0.02 & 18.11 $\pm$ 0.005 & 25.94 & 25.78 & 24.82 & 0.16 $\pm$ 0.03 \\
 {\it CORE}       & 166  & 11.39  & 22.21 $\pm$ 0.05 & 22.13 $\pm$ 0.06 & 22.22 $\pm$ 0.03 & 24.85 & 24.77 & 24.23 & 0.08 $\pm$ 0.11 \\ 
TOTAL & 7203 & 494.44 & 19.16 $\pm$ 0.01 & 19.01 $\pm$ 0.02 & 18.07 $\pm$ 0.005 & 25.90 & 25.74 & 24.80 & 0.15 $\pm$ 0.03 \\
\hline                                   
\end{tabular}
\end{table*}

Photometry of the stream was also obtained in the $u$'-band using the TNG data of NGC~7241 described in Sec.~\ref{sec:2.2}. and the same elliptical aperture and masking as for the bands $g$ and $r$. To find the zeropoint of the u-band image ($30.18\pm0.04$ mag), matched aperture photometry was used in comparison with {\it Sloan Digital Sky Survey} (SDSS) images. With that purpose, we ran {\sc NoiseChisel} and Segment and selected "Clumps" with S/N > 10 and axis-ratio > 0.85 in the u-band image. An aperture of radius 2 arcsec was then placed over them in the TNG and SDSS images (using {\sc MakeProfiles}) and the photometry over the apertures obtained with {\sc MakeCatalog}. The zeropoint was estimated by calculating the difference (for stars with magnitude between 17 to 20). 

The $u$'-band photometry  is also included in Tab.~\ref{tab:1} and discussed in Sec.~\ref{sec:4.3}. Finally, the width of the stellar stream has been estimated for three points along the stream (maximum, minimum and intermediate width) assuming a distance of 22.5kpc. The results are given in Tab.~\ref{tab:2}.

Despite the low exposure times of the GALEX images, the emission from the stellar stream is also detected in both bands due certainly to its very blue color. Using circular apertures of 15 arcsec ($\sim$3 times the FWHM of the GALEX PSF) in radius centered on RA(J000) = 22:15:52.38 and Dec(J2000) = 19:11:29.3 we estimated FUV and NUV aperture magnitudes of 20.10\,mag and 20.06\,mag with estimated errors between 0.05-0.11\,mag and 0.16-0.22\,mag,
respectively. The FUV-NUV color inferred is (FUV-NUV)=0.045 with an error no lower than 0.18\,mag. The UV photometry was performed as in Gil de Paz et al$.$ (2007), so it takes into account the highly Poissonian regime of the GALEX AIS images and the potential presence of low-frequency background variations. The blue color found corresponds to a
roughly flat $f_{\nu}$ spectrum is comparable to that measured in
star-forming dwarfs and the outermost regions of spiral disks \citep[see ][]{2018ApJS..234...18B}, including Extended Ultra-Violet disks (XUV-disk), and it is commonly associated with regions of active recent star formation and low dust content.

\subsection{Metallicity }

We estimate the metal abundance of the ISM in the stream of NGC~7241 using the two emission lines clearly
detected in our optical spectra, namely [\ion{O}{iii}]$\lambda$5007\,\AA\ and
H$\beta$. The best-fitting gaussians to each of these two lines after
adding the spectra of the 11 spaxels where line emission is clearly
detected ($\mathrm{S/N}\gtrsim5$ at the peak of the line) yield
[\ion{O}{iii}]$\lambda$5007/H$\beta$ line ratio of
$-$0.23$^{+0.14}_{-0.21}$. Alternatively, the direct sums of the fluxes
above the continuum in the (rest-frame) range between 4855\,\AA\ and
4867\,\AA\ yield a ratio of $-0.09^{+0.16}_{-0.27}$. Based exclusively
on this single-line ratio we cannot obtain too much information about the ionization
properties in the possible progenitor of the stream {\it STREAM-CORE} (either on the conditions of the gas or the
ionizing spectrum). This line ratio, according to the typical ones measured in HII regions in the normal galaxies included in the CALIFA sample \citep[see  top-left panel in Figure 7 of ][]{morisset2016},  favors an Oxygen abundance most likely in the range 8.35 $<$ 12 + log(O/H) $<$ 8.55 (see their colorbar), or between 0.4$\times$Z$_{\odot}$ and 0.63$\times$Z$_{\odot}$ \citep[adopting a solar oxygen abundance value of logA(O)=8.75;][]{10.1093/mnras/stab2160}.

\section{Discussion}

\subsection{The NGC~7241 and its dwarf companion}\label{sec:hoststream}

In agreement with the footprints of an NGC~7241 companion in the HI distribution \citep{Giovanelli1984} and the final confirmation by \citet{Leaman2015}, we show that the dwarf irregular that overlaps the main body of NGC~7241 is clearly visible in the Mount Lemmon 0.80-meter deep image  (Fig.~\ref{fig:1}, bottom panel). We named this new system {\it DWARF-COMPANION}, hereafter.\\
For the superimposed {\it DWARF-COMPANION} object (see Fig.~\ref{fig:1}, top-right panel) in \citet{Leaman2015} the authors estimate a $M_{*} \sim 1.5\times10^{8}$M$_{\odot}$, also obtained from SED fitting. Though very uncertain, the size and H$\alpha$ velocity dispersion of the {\it DWARF-COMPANION} imply a dynamical mass of order $\sim 10^{8}$M$_{\odot}$, so this would be a baryon dominated object.\\
Using this data on the total mass of each galactic system, in the following sections we will focus on the origin and properties of the blue stream detected in the NGC~7241 and its dwarf companion's neighbourhood.

\subsection{Who is the progenitor of the blue stellar stream?} \label{sec:progenitor}

As mentioned in Sec.~\ref{sec:radvel}, the blue stream around NGC~7241 seems to emanate from a bright compact object we named {\it GC-CAND}, that is close to another bright but fainter structure we called {\it STREAM-CORE}. In this section we study which of these two objects is the best candidate to be the stream's progenitor.\\
{\it GC-CAND} (R.A.=22:15:52.4  Dec.=+19:11:32.9) is an object that was also catalogued in the Gaia early DR3  \citep{2021AA...649A...1G}. It has a Gaia G magnitude of G=20.2 with the assumed distance this comes out to absolute G magnitude of $M_{\rm G}=-11.5$ which roughly corresponds to $M_{\rm V}=-11$. This absolute magnitude places this source in the very bright end of Globular Clusters ($M_{\rm V}<-10$) where we know that more than 50\% of sources are the relic nuclear star cluster of the galaxy that is being stripped \citep{Voggel2019}. Streams with such embedded former nuclei have been discovered in other galaxies \citep[e.g.][]{Jennings2015}. {\it GC-CAND} also has a larger Gaia BP-RP excess factors similar to what is seen for GCs/Nuclei in nearby galaxies \citep{Voggel2020}. Gaia proper motions have large errors at these faint magnitudes and its proper motion in RA direction is detected at $2.5\,\sigma$ level, which is not significant enough to clearly mark it as foreground star. Therefore, in order to confirm whether the {\it GC-CAND} compact object is associated to the stream, a radial velocity measurement is required.  In Sec.~\ref{sec:radvel} we show that we obtain a velocity of $V_h=-233\pm33$~km\,s$^{-1}$ with respect to the solar system barycenter, therefore, this compact object is not associated with NGC~7241 that has a systemic velocity  of about 1500~km\,s$^{-1}$ but an accidental foreground Milky Way star. To explore this further, we make a simple estimate of the velocity range expected in this line-of-sight using the SDSS DR16 spectra archive \citep{2020ApJS..249....3A}. Within a 30~arcmin cone around the object, there are $\sim$ 50 stars with observed SDSS spectra. Their heliocentric velocities range from $-560$ to $+125$ with a median value of $-54$ and a standard deviation of $89$~km\,s$^{-1}$, estimated by the median absolute deviation. The star velocity distribution shows the second peak near $-200$~km\,s$^{-1}$  and 8 of 50 stars have velocity less than $-230$~km\,s$^{-1}$. Thus, we conclude that {\it GC-CAND} has a radial velocity which is typical for Galactic halo stars and cannot be the core of this stellar stream.\\

The second bright source we detected close to the stream's geometrical center is the {\it STREAM-CORE}. From our broad-band photometry (see Tab.~\ref{tab:1}) we obtained that it would have an absolute magnitude of M$_{g}\sim$ -10  at the distance of NGC~7241. We also find that its color is  bluer than that of the stream, as expected from the emission only observed in this region in the MEGARA spectra (see discussion in Sec.~\ref{sec:radvel} and Fig.~\ref{fig:2}, center panel).  In Sec.~\ref{sec:radvel} we showed that the systemic radial velocity of {\it STREAM-CORE} region is 1547.92$\pm$1.65\,km\,s$^{-1}$ (from H$\beta$ in emission), this value is similar to the one of both, the NGC~7241 and its {\it DWARF-COMPANION}, so it is not a strong statement to assume that the {\it STREAM-CORE} is somehow related with those systems. In addition, the shape of the stream around the position of the {\it STREAM-CORE} (see Fig.~\ref{fig:3}, upper panel) resembles the typical structure of a tidally stripped dwarf galaxy \citep[see e.g.,][]{Wang22}. Its location in the geometrical center of the stream, radial velocity, color, and shape allows us to conclude that this over-density is likely the actual progenitor of the stream, which is undergoing a current episode of induced star formation by the tidal interaction with the NGC~7241 or its foreground dwarf companion. Obviously, the presence of gas in the progenitor of the stream is essential to form new stars, this suggest {\it STREAM-CORE} cannot be a typical globular cluster, since those have no gas, but probably the main body of an accreted dwarf satellite. 

\begin{table}
\centering                          
\caption{Minimum, maximum and median estimations of the NGC~7241's stellar stream on-sky width measurements. We include an estimation of its physical width assuming a distance to the stream of 22.4~Mpc \citep{Leaman2015}.} 
\label{tab:2}     
\begin{tabular}{c c c c c}        
\hline\hline                 
       & RA    & DEC      & Width & Width \\
       &(deg)     & (deg)      & (arcsec) & (kpc) \\
\hline                        
   Max.      & 333.97123 & 19.18.6971 & 11.960 & 0.80 \\ 
   Min.      & 333.96934 & 19.190611  & 7.372 & 1.30 \\
   Med. & 333.97307 & 19.183051  & 8.798 & 0.95  \\ 
\hline                                   
\end{tabular}
\end{table}

\subsection{Properties of the blue stream's progenitor}\label{sec:4.3}

In order to better understand the NGC~7241 stream, we can make a quantitative estimate of which should have been the NGC~7241 stream progenitor's mass using the results of \cite{johnston2001} and \cite{erkal2016} who derived mass estimates for stream progenitors assuming that the stream plane is viewed face-on or edge-on, respectively. These estimates work by determining how the orbits of stars in a given potential fan out once they are ejected from the progenitor and connecting this to the progenitor's dynamical mass.\\
First we will make the computations assuming that the stream is associated with the central galaxy, not with the dwarf and is seen face-on.  Assuming that the stream is on a circular orbit at $20$ kpc and that the circular velocity at this radius using the mass estimations in Sec.~\ref{sec:hoststream}, is 202~km s$^{-1}$, we can use Eq. 12 in \cite{johnston2001} to obtain that the progenitor mass of the stream is between $1.3 - 5.5\times10^7 M_\odot$ using the minimal and maximal width of the stream (see Tab.~\ref{tab:2}), respectively. We note that this mass estimate is linearly proportional to the stream's pericenter and thus the progenitor mass could be substantially lower if the stream is on an eccentric orbit at apocenter at present day. Alternatively, if we now assume that we are observing the stream edge-on, we can use the results of \citet{erkal2016} (Eq. 23) and we obtain a progenitor mass of $6.2\times10^7 - 2.7\times10^8 M_\odot$ for the minimal and maximum stream width (Tab.~\ref{tab:2}), respectively.  We note that these mass estimates are much lower than for a stream with the same width in the Milky Way since NGC~7241 is a much less massive galaxy, which yields a larger tidal radius (i.e., width) for the stream presented in this work.\\ Similarly, we can obtain the progenitor's mass assuming that the stream is linked to the {\it DWARF-COMPANION} and not to the central galaxy. Using the \cite{johnston2001} approach (its Eq. 12) and the {\it DWARF-COMPANION} mass estimations obtained in Sec.~\ref{sec:hoststream} we find that the progenitor mass of the stream should be as low as $3.2\times10^4 - 1.4\times10^5 M_\odot$.

We note that these estimates suggest that if the stream is on a near circular orbit around NGC~7241, its progenitor is likely a dwarf galaxy stream. Otherwise, if it is on a very radial orbit pointing towards NGC~7241, or it is orbiting the {\it DWARF-COMPANION} then it could come from a lower mass system like a globular cluster.

Interestingly, the stream width increases with projected distance from NGC~7241/{\it DWARF-COMPANION}  (see e.g., Fig.~\ref{fig:1}). If the stream is on a very eccentric orbit, this may be due to the characteristic shell-like debris which causes the stream width to increase with radius \citep[e.g. see Fig. 8 of ][]{hendel2015}. Alternatively, if there is a strong distance gradient along the stream, where the parts of the stream that are closest to the center of NGC~7241/{\it DWARF-COMPANION} in projection is in fact farthest away from us in 3D, it is possible that the most distant debris is too diffuse to observe. The change in the apparent width of the stream could, in that case, merely be an observational effect. If the stream plane is instead being viewed close to edge-on, the change in the stream width could be due to an overlap with different parts of the stream along the line of sight. The specific viewing angle can have a large impact on how we interpret tidal debris \citep[e.g.,][]{barnes2009}, and can lead to apparent asymmetric observations of intrinsically symmetric tidal features. Lastly, the stream could physically be asymmetric in nature.  Recent data from the Milky Way have revealed that several stellar streams are asymmetric (e.g., Jhelum: \citealt{bonaca2019} and Pal 5: \citealt{bernard2016,bonaca2020}). Asymmetric stellar streams can be produced through various mechanisms such as interactions with galactic bars  \citep[][]{price2016,erkal2017, pearson2017}, heating of parts of the streams due to interactions with dark substructure  \citep[e.g.,][]{johnston2002,ibata2002,bonaca2014}, interactions with nearby systems, or specific orbit families within the parent potential causing parts of the stream to ``fan'' out  \citep[e.g.,][]{pearson2015,fardal2015,price-whelan2016, yavetz2021}. It is unclear from the edge-on view of NGC~7241 whether the galaxy hosts a bar, as we do not observer any evidence of an x-shaped central structure typically found in barred spirals \citep[e.g.,][]{bureau2006}, but it is clear that the stream could be perturbed by the presence of the second object in the system, NGC~7241 or {\it DWARF-COMPANION}. Additionally, if interaction with the bar was the cause, the stream would have already needed to have a close pass with the center of NGC~7241 but no evidences of such interaction are present in the system.\\
Deeper data in the future will hopefully allow us to distinguish between the possible scenarios discussed in this section.

\subsection{Is NGC~7241 the host galaxy of the blue stellar stream?}\label{sec:ishoststream}

With the mass estimations presented in Sec.~\ref{sec:hoststream} in hand, if the stellar stream has been generated by the tidal interaction with NGC~7241 and has a mass of $M_{\rm stream} \sim 2\times10^{7}$~M$_{\odot}$ (Sec.~\ref{sec:4.3}), then its tidal radius would be $R_{\rm tid} \sim 823$ pc or $\sim 8^{\rm ‘’}$. Conversely, the tidal radius for the object with respect to the potential of the baryon dominated {\it DWARF-COMPANION} with a mass $M_{\rm stream} \sim 1\times10^{5}$~M$_{\odot}$ would be $R_{\rm tid} \sim 600$ pc or $\sim6^{\rm ‘’}$.\\
In the previous sections we showed that the width of the stream ranges from 800 pc to 1.3 kpc. These values are larger than the tidal radii crudely estimated from the mass profile and projected distance of NGC~7241, and also than the one derived from the companion dwarfs' mass, so it would seem sensible to assume that the disrupting system can be dominated by either the tidal influences from the potential of NGC~7241, or that from the dwarf galaxy companion alone. This result is highly sensitive on the original mass of the stream's progenitor, a value that is highly uncertain (see Sec.~\ref{sec:4.3}) so we can not use this argument alone to reach any conclusion.\\
Is there any other evidence that supports the hypothesis that the stream have been bound to the SMC-like companion dwarf ({\it DWARF-COMPANION}) prior to falling into NGC~7241?\\  The apparently young dynamical age of the stream \citep[see Section 6.4 in ][]{Leaman2015}, and the similar velocity offset \citep[$\sim 190 ~{\rm km/s}$][]{Leaman2015} of the stream and {\it DWARF-COMPANION} with respect to the NGC~7241 would seem to favour this scenario.  Additionally, if the dwarf irregular satellite is undergoing a relatively recent infall towards the NGC~7241 (see further discussion in Sec.~\ref{sec:4.4}), and thus it still keeps its own substructure (the stream) it may explain why it and the central galaxy seem to show signatures of a recent star formation burst.  Whether or not it is expected that an SMC-mass system like the {\it DWARF-COMPANION} has a companion dwarf of the mass of the stream progenitor is discussed further in Sec.~\ref{sec:4.4}.\\
In conclusion, with the current data, we can not confirm that the stream is associated with the dwarf companion, although none of the analysis presented here disfavours this hypothesis.

\subsection{The origin of blue streams: insights from theory, simulations and observations of the local group dwarf's SFH}\label{sec:4.4}
\begin{figure*}
	\includegraphics[width=\textwidth]{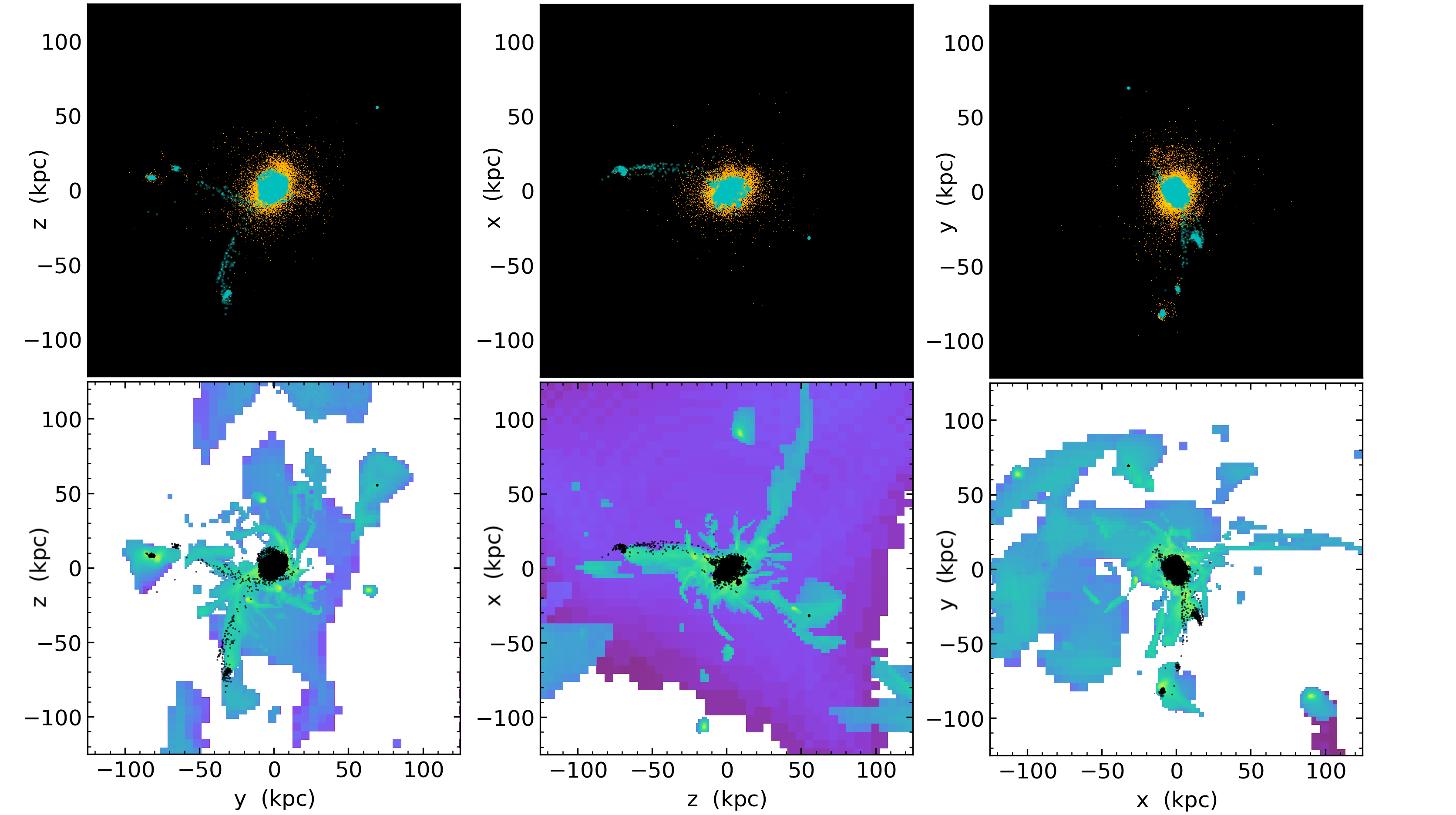}
    \caption{Blue streams in the GARROTXA cosmological simulation of a z=0 MW-mass galaxy, observed at z=1.5. {\it Top panels}: Total stellar surface luminosity. {\it Bottom panels}: Projected cold gas density (T $<$ 5$\times$10$^4$~K){\bf , green-yellow are high-density gas regions, blue-black are low-density}. Cyan dots (left) and black dots (right) show the position of the star particles younger than 350~Myr.}
    \label{fig:Simulations}
\end{figure*}

Star-formation is rarely detected in stellar streams, as it requires the presence of dense molecular clouds that would be quickly heated up and destroyed by the many processes involved in the satellite-central galaxy interaction \citep[e.g.][]{Emerick2016, 2021PASA...38...35C}. As a consequence, most tidal streams in the local universe host old stars that were stripped out from its progenitor's dwarf galaxy by a minor merger event. However, this scenario does not apply to all dwarf-central galaxy interactions. From the theory of galaxy formation and evolution we expect that changes on the mass of the central galaxy modify the efficiency that the gas ram pressure stripping process has on removing gas from the incoming satellites \citep{Koutsouridou2019}. These changes probably occurred in the Milky Way - Andromeda galactic systems when transiting from a low-mass system with no warm-hot corona to the current situation with a well defined warm-hot CGM.

In the early scenario where dwarf galaxies merged with a low-mass MW with no well developed warm-hot CGM, the infalling satellites were able to keep their gas for a long time, and thus, we expect they could suffer a strong star formation burst \citep{DiCintio2021} similar to the one observed in the NGC7241's companion, or in observations of similar interacting systems \citep[e.g.][]{Kapferer2009,Beaton2014}. This star formation peak would be clearly reflected in their SFH. Many authors obtained the SFH of the local group dwarfs by fitting the color-magnitude diagrams to evolutionary models and reported an SFH with a single peak \citep{Aparicio2009,Weisz2014}, and obtained results that may be consistent with the aforementioned hypothesis and with what is currently occurring in the NGC~7241 system. However the exact shape of the SFH of these objects is difficult to estimate as it depends on multiple variables, for example, more massive and concentrated satellites are more efficient on shielding gas in their central regions from the ram pressure stripping, so they can keep forming stars for longer times \citep{2021PASA...38...35C, Font2021}. Although this degeneracy is present, a gas rich dwarf galaxy will always suffer a strong star formation burst in its first interaction with the central galaxy due to compression of molecular clouds induced by tidal forces \citep{DiCintio2021}. In this scenario it is expected that the gas clouds and stars in the star forming complexes generated as a consequence of the interaction will be affected by the tidal forces from the central system and will became part of the resulting tidal streams. Therefore, this new stream, at least in the first orbit, will include young stars and gas that has not been yet completely stripped out from the progenitor, i.e., will be a show up as a blue stream. \\
Recent zoom-in cosmological simulations captured this process at high-z for z=0 MW-mass systems, i.e., when these were low-mass enough, showing the formation of blue streams \citep{RocaFabrega2016,Buck2019}\footnote{In the context of cosmological simulations it is important to mention that this scenario is only weakly dependent on redshift at z$<$1.5 and that results at z$\sim$1 can be easily extrapolated to z=0. This because the minor merger rate is only slowly decreasing at this z \citep{Lotz2011} and because, although the UV background radiation decays fast with z \citep{Faucher2020}, this has no strong impact on the dense star-forming regions which are partially or completely self-shielded. Additionally, this weak dependence with z is clearly exceeded by the central galaxy mass dependence, as shown above. Thus, we expect that blue streams around low-mass central galaxies exist from high-z (z$\sim$1.5) down to the local Universe, as confirmed by cosmological simulations.}. An example is the GARROTXA model \citep{RocaFabrega2016}, a high-resolution simulation of the formation of an MW-mass galactic system. In this model, we detected that at $z\sim1.0-1.5$, that is when the progenitor has a mass of $\sim$10$^{11}$M$_{\odot}$, satellites almost reach the galactic disk region without losing their cold gas (see Fig.\ref{fig:Simulations}, bottom panels). We also observe ongoing strong star formation inside the satellite's core and also on the stripped stellar stream, i.e., we observe the formation of a blue stream (see Fig.\ref{fig:Simulations}, top panels). Interestingly, satellites contain almost no stars previous to their interaction with the MW-mass progenitor, therefore this interaction induces the first strong star formation episode within them. In this same model we also observed that at z=0, i.e. when the MW-mass system already developed a warm-hot gas corona \citep{RocaFabrega2016}, satellites with a similar initial mass and gas content cannot retain their cold gas and show no star formation in the tidal streams.\\
These results suggest that in the observed blue stream we may be watching a tidally induced star formation burst equivalent to the one that shaped the SFH of some of the dwarf galactic systems in the Local Group \citep{Weisz2015, DiCintio2021} when the MW and Andromeda were low mass galaxies.\\

Before finishing the discussion section we want to emphasise that we cannot discard the possibility the blue stream is associated with the star-forming dwarf galaxy ({\it DWARF-COMPANION}), not with the NGC~7241 itself. In recent years there has been a growing interest in detecting the presence of streams around low-mass galactic systems, i.e., LMC and SMC-like galaxies \citep[e.g.,][]{MartinezDelgado2012,KadoFong2020}. The $\Lambda$CDM theory of a hierarchical universe predicts the presence of very low-mass halos interacting with dwarf galaxies in low-density environments \citep[e.g.,][]{Guo2011,Sales2013}. Some examples of these interactions were recently discovered by the MADCASH Survey \citep{Carlin2016,Carlin2019}. In this context, we propose an alternative scenario: a small galactic system interacted with the {\it DWARF-COMPANION} inducing the observed star formation burst and generating the observed blue stream. Later, both of them fell simultaneously towards the NGC~7241. Metallicity estimations for each one of the components in the system could help to unveil which is the true scenario. However, the oxygen abundances we obtained here are too uncertain to allow us to confirm or discard any hypothesis. So, we conclude that more observations are needed to unveil the real scenario posed by this complex galactic system.

\section{Conclusions}

In this paper,  we have characterized the structure of the blue star-forming stream of NGC 7241 discovered by \citet{Leaman2015}, and the kinematics of its most plausible progenitor ({\it STREAM-CORE}) using new available images in different broad-bands from four independent sources. In order to explore if this stream had a globular cluster-like progenitor, we obtained follow-up observations using the MEGARA instrument at the GTC in its IFU configuration.

We concluded that the  bright compact object ({\it GC-CAND}) embedded in the stream that we initially considered  as the most probable progenitor is actually a foreground halo star overlapping the path of the stream. However, we report  the discovery of a fainter blob in the stream displaying emission lines ({\it STREAM-CORE}) which could be interpreted as the actual progenitor displaying a burst of star formation. Our photometric study suggests that this on-going star formation region detected within the stream produces a near-UV luminosity  comparable with those observed in star-forming dwarfs and the outermost regions of spiral disks. We also measured the width of the stream and obtained values in the range of the observed tidal streams around the Milky Way. Based on these width estimate, we constrain that the progenitor's mass, if the stream was created by tidal interaction with NGC~7241, is between 6.4$\times$10$^{6}$~M$_{\odot}$ and 2.7$\times$10$^{8}$~M$_{\odot}$ when assuming the in-fall was is in a circular orbit. For a more radial orbit, of if it was created by tidal interaction with the ({\it DWARF-COMPANION}) this estimate would result in a much lower mass even compatible with the one of a massive globular cluster. Although deeper data is needed to distinguish between these possibilities,  the NGC~7241 stream is the stream with the lowest mass progenitor reported so far, providing a lower limit to the detection of narrow stellar streams beyond the Local Group \citep[see also][]{pearson2019,pearson2021}.

Blue star-forming streams like that observed around NGC~7241 should be a common occurrence around galaxies of similar mass. The current galaxy formation and evolution theories predict the presence of these structures around disk galaxies with stellar masses below or around 10$^{10.5}$~M$_{\odot}$ whose have not developed a warm-hot CGM and, therefore, the merging dwarf galaxies can keep their gas for longer times. We show that these structures are also commonly observed in high-resolution cosmological simulations of galactic systems with similar mass as NGC~7241. This scenario also agrees well with the tidal interaction and ram pressure stripping processes that produced the observed bursty star formation history on some of the local group dwarf galaxies followed by a fast quenching \citep[e.g.,][]{Kapferer2009,DiCintio2021} when they first entered the virial radius of Andromeda and the MW progenitors \citep{Aparicio2009,Weisz2014}.

\section*{Acknowledgements}
We thanks to Tobias Buck for a fruitful discussion about blue stellar streams in cosmological simulations. We also thanks to the referee for very constructive comments which helped to improve this manuscript. DMD acknowledges financial support from the Talentia Senior Program (through the incentive ASE-136) from Secretar\'\i a General de  Universidades, Investigaci\'{o}n y Tecnolog\'\i a, de la Junta de Andaluc\'\i a. DMD acknowledge funding from the State Agency for Research of the Spanish MCIU through the ``Center of Excellence Severo Ochoa" award to the Instituto de Astrof{\'i}sica de Andaluc{\'i}a (SEV-2017-0709) and project (PDI2020-114581GB-C21/ AEI / 10.13039/501100011033). SRF acknowledge financial support from the Spanish Ministry of Economy and Competitiveness (MINECO) under grant number AYA2016-75808-R, AYA2017-90589-REDT and S2018/NMT-429, and from the CAM-UCM under grant number PR65/19-22462. SRF acknowledges support from a Spanish postdoctoral fellowship, under grant number 2017-T2/TIC-5592. MAGF acknowledges financial support from the MICINN project PID2020-114581GB-C22 grant (Spain).  MA acknowledge the financial support from the European Union - NextGenerationEU and the Spanish Ministry of Science and Innovation through the Recovery and Resilience Facility project J-CAVA. JG thanks support from the project AYA2018-RTI-096188-b-i00. JIP acknowledges finantial support from projects Estallidos6 AYA2016-79724-C4 (Spanish Ministerio de Econom\'{\i}a y Competitividad), Estallidos7 PID2019-107408GB-C44 (Spanish Ministerio de Ciencia e Innovaci\'{o}n) and grant P18-FR-2664 (Junta de Andaluc\'{\i}a).
 
 Based on observations made with the Gran Telescopio Canarias (GTC), installed at the Spanish Observatorio del Roque de los Muchachos of the Instituto de Astrofísica de Canarias, in the island of La Palma. MEGARA has been built by a Consortium led by the Universidad Complutense de Madrid (Spain) and that also includes the Instituto de Astrofísica, Óptica y Electrónica (Mexico), Instituto de Astrofísica de Andalucía (CSIC, Spain) and the Universidad Politécnica de Madrid (Spain). MEGARA is funded by the Consortium institutions, GRANTECAN S.A. and European Regional Development Funds (ERDF), through Programa Operativo Canarias FEDER 2014-2020. This publication is based on observations made on the island of La Palma with the Italian Telescopio Nazionale {\it Galileo}, which is
operated by the Fundaci\'on Galileo Galilei-INAF (Istituto Nazionaledi Astrofisica) and is located in the Spanish Observatorio of the Roque de Los Muchachos of the Instituto de Astrof\'isica de Canarias. 

This project used public archival data from the {\it DESI Legacy Imaging Surveys} (DESI LIS). The Legacy Surveys consist of three individual and complementary projects: the Dark Energy Camera Legacy Survey (DECaLS; Proposal ID 2014B-0404; PIs: David Schlegel and Arjun Dey), the Beijing-Arizona Sky Survey (BASS; NOAO Prop. ID 2015A-0801; PIs: Zhou Xu and Xiaohui Fan), and the Mayall z-band Legacy Survey (MzLS; Prop. ID 2016A-0453; PI: Arjun Dey). DECaLS, BASS and MzLS together include data obtained, respectively, at the Blanco telescope, Cerro Tololo Inter-American Observatory, NSF’s NOIRLab; the Bok telescope, Steward Observatory, University of Arizona; and the Mayall telescope, Kitt Peak National Observatory, NOIRLab. The Legacy Surveys project is honored to be permitted to conduct astronomical research on Iolkam Du’ag (Kitt Peak), a mountain with particular significance to the Tohono O’odham Nation. NOIRLab is operated by the Association of Universities for Research in Astronomy (AURA) under a cooperative agreement with the National Science Foundation. This project used data obtained with the Dark Energy Camera (DECam), which was constructed by the Dark Energy Survey (DES) collaboration. Funding for the DES Projects has been provided by the U.S. Department of Energy, the U.S. National Science Foundation, the Ministry of Science and Education of Spain, the Science and Technology Facilities Council of the United Kingdom, the Higher Education Funding Council for England, the National Center for Supercomputing Applications at the University of Illinois at Urbana-Champaign, the Kavli Institute of Cosmological Physics at the University of Chicago, Center for Cosmology and Astro-Particle Physics at the Ohio State University, the Mitchell Institute for Fundamental Physics and Astronomy at Texas A\&M University, Financiadora de Estudos e Projetos, Fundacao Carlos Chagas Filho de Amparo, Financiadora de Estudos e Projetos, Fundacao Carlos Chagas Filho de Amparo a Pesquisa do Estado do Rio de Janeiro, Conselho Nacional de Desenvolvimento Cientifico e Tecnologico and the Ministerio da Ciencia, Tecnologia e Inovacao, the Deutsche Forschungsgemeinschaft and the Collaborating Institutions in the Dark Energy Survey. The Collaborating Institutions are Argonne National Laboratory, the University of California at Santa Cruz, the University of Cambridge, Centro de Investigaciones Energeticas, Medioambientales y Tecnologicas-Madrid, the University of Chicago, University College London, the DES-Brazil Consortium, the University of Edinburgh, the Eidgenossische Technische Hochschule (ETH) Zurich, Fermi National Accelerator Laboratory, the University of Illinois at Urbana-Champaign, the Institut de Ciencies de l’Espai (IEEC/CSIC), the Institut de Fisica d’Altes Energies, Lawrence Berkeley National Laboratory, the Ludwig Maximilians Universitat Munchen and the associated Excellence Cluster Universe, the University of Michigan, NSF’s NOIRLab, the University of Nottingham, the Ohio State University, the University of Pennsylvania, the University of Portsmouth, SLAC National Accelerator Laboratory, Stanford University, the University of Sussex, and Texas A\&M University.
The Legacy Surveys imaging of the DESI footprint is supported by the Director, Office of Science, Office of High Energy Physics of the U.S. Department of Energy under Contract No. DE-AC02-05CH1123, by the National Energy Research Scientific Computing Center, a DOE Office of Science User Facility under the same contract; and by the U.S. National Science Foundation, Division of Astronomical Sciences under Contract No. AST-0950945 to NOAO.

Based in part on observations at Cerro Tololo Inter-American Observatory, National Optical Astronomy Observatory, which is operated by the Association of Universities for Research in Astronomy (AURA) under a cooperative agreement with the National Science Foundation.

Support for this work was provided by NASA through the NASA Hubble Fellowship grant \#HST-HF2-51466.001-A awarded by the Space Telescope Science Institute, which is operated by the Association of Universities for Research in Astronomy, Incorporated, under NASA contract NAS5-26555.

\section*{Data Availability}

The data underlying this article will be shared on reasonable request to the corresponding author.



\bibliography{ref} 






\label{lastpage}
\end{document}